\documentclass{article}
\usepackage{amssymb}

\usepackage{graphicx}
\usepackage{amsmath}
\usepackage{epsfig}
\usepackage{color}

\begin{document}

	\title{Climbing plants --
		Wrapping elastic plant stems around a cylindrical stake}

	\author{ Henri Gouin \thanks{   
			E-mails:
			henri.gouin@univ-amu.fr; henri.gouin@ens-lyon.org} }
	\date{\footnotesize Aix--Marseille University,
		IUSTI,  CNRS UMR 7343, Marseille, France.}
	\maketitle

%	\date{march 2025}
	
	% The correct dates will be entered by the editor
	\maketitle
 
\begin{abstract}
	 Since Charles Darwin's time, the study of climbing plants on a cylindrical stake has been the subject of numerous articles in plant biology.  One of the main ideas for studying the coiling of an elastic plant stem is to consider the growth of the plant stem in terms of evolution over time. However, as this development takes place over a long time scale, the static study alone has not been studied independently. Our static approach requires us to take into account elasticity, turgor pressure and gravity forces in a first analysis.\\
	 The aim of this article is to present a simplified model demonstrating why plant stems climb mainly on their circular helix-shaped stakes, with the diameter of the stake playing an important role in plant stem ascent, as does the fineness of the stem. To perform this calculation, for a given mass density, we consider the variational principle of minimum energy. For thin plant stems,
	 we can see, in first approximation, that  the effect of gravity and turgor pressure can be neglected with respect to the energy of elasticity, and that the bulk of the calculation concerns elasticity terms.
\end{abstract}
\section{Introduction} 
\subsection{About climbing plants}
Climbing plants use a variety of means to reach the light and can transform a space by adding greenery and life, while requiring a certain degree of attention to staking.  Voluble plants, such as certain varieties of wisteria, wrap their stems around a support. 
Work on climbing plants and their association with guardians covers a wide range of disciplines. The interaction between climbing plants and their supports is complex and multidimensional; it involves physiological mechanisms of growth and attachment, responses to environmental stimuli and ecological implications. Understanding these interactions is essential to optimize their cultivation.
\\
Some plants develop stems that adopt a helical shape to maximize their exposure to light while wrapping around supports. This shape enables them to grow efficiently. This behavior is linked to the need to support the plant while minimizing the energy expended for growth. Many climbing plants, such as sweet peas and certain varieties of liana, have been observed to have stems that grow in a helix. Some climbers, such as climbing roses, may have relatively straight stems. These stems can extend vertically without twisting, enabling them to reach the light quickly. Other plants, such as certain lianas, can develop stems with well-defined angles. This angular geometry may be an adaptive strategy, enabling plants to move in response to obstacles and
variations in the shape of climbing plant stems, whether straight, angular or helical, are the result of a complex combination of ecological, environmental and biological factors.
\subsection{Study of climbing plants associated with elasticity}
The biological growth of plant stems is a fascinating process of great complexity that has attracted the attention of generations of biologists and today  remains questions associated with elasticity \cite{goriely_2008}.
First of all, Lockhart points out that the time scale for reaching elastic equilibrium in plants is much faster than the time scale associated with the  extension of stems.
As it grows, the plant stem system must therefore remain in a state very close to that of the elastic equilibrium  \cite{Lockhart_1965}.
\\
We study an elastic plant stem that wraps around a rigid cylinder with a circular base \cite{Darwin_1875,Gianoli_2015,Goriely_2017,Goriely_Neukirch_2006,Guo_2019,Isnard_Silk_2009,Neukirch_2009}. The plant's stem is a small rod, bent and twisted so that it wraps around the rigid cylinder without clinging to it.\\
We assume that the ascent of the rod on its support satisfies the following two conditions corresponding to an extremum of length and an extremum of energy as it is done in \cite{Gouin_2020}   

a) The plant stem connects two points of the vertical stake along an extremal  (minimum) length on its support.

b) The energy of the plant stem, which is the sum of the elastic deformation energy, the energy due to gravity forces and the turgidity energy of the plant shoot, is assumed to be extrema  (minimum).  \\
It is supposed that the plant stem has no point attaching it to the vertical post.
In the reference space $\mathcal D_0$ of the plant stem, the centers of the circles delimiting the straight sections of the small cylinder constituting the stem form a line segment denoted $(\it\Gamma_0)$ carried by the axis $\overrightarrow{\boldsymbol k}$ of the space, which is also the axis of the cylindrical stake \cite{Berdichevskya_2009,Gouin_2007}. Let's denote $(\it\Gamma)$ the curve of the plant stem, image of $(\it\Gamma_0)$ in actual physical--space $\mathcal D$.
\\
The $(\it\Gamma)$--curve is assumed to admit a Frenet  frame denoted $M\, \boldsymbol {\overrightarrow{t}, \overrightarrow{n}, \overrightarrow{b}}$ in $\mathcal D$. This Frenet  frame is assumed to be image of $M_0\, \boldsymbol {\overrightarrow{i}, \overrightarrow{j}, \overrightarrow{k}}$ as  the frame in the reference space $\mathcal D_0$, where $M_0$ is the current point of $(\it\Gamma_0)$, and $M$ its image on the $(\it\Gamma)$--curve. The distance from $M$ to the $O\boldsymbol{\overrightarrow k}$ axis is denoted $R_1$ (see Figure 1).
\\
Since circular cylinders are developable surfaces, the planar development of the $(\it\Gamma)$--curve is a straight line segment. As a result, the plant stem deformed by coiling on the stake is such that the $(\it\Gamma)$--curve, which must be of minimum length, is a circular helix of axis $O\boldsymbol{\overrightarrow k}$.\\
The displacements of  Frenet  frame $M\, \boldsymbol {\overrightarrow{t}, \overrightarrow{n}, \overrightarrow{b}}$ of the $(\it\Gamma)$--curve  taken relative to the fixed orthonormal direct frame $\displaystyle O\, \boldsymbol {\overrightarrow{t}, \overrightarrow{n}, \overrightarrow{b}}$ are called $d\boldsymbol {\overrightarrow t}, d\boldsymbol {\overrightarrow n}, d\boldsymbol {\overrightarrow b}$. By their very nature, these are solid displacements. The displacement of  origin $M$ of Frenet's reference frame is denoted $d\boldsymbol {\overrightarrow M} \equiv d{\overrightarrow {OM}}$.\\
Frenet's relations are used to obtain the displacements of   plant stems. They are expressed as a function of a parameter denoted $s$ representing the curvilinear abscissa of $(\it \Gamma)$. The result is as follows \cite{Guggenheimer_1977,Kobayashi_Nomizu_1996}
\begin{equation}
\left\{
\begin{array}{l}
 d\boldsymbol {\overrightarrow t} = \displaystyle\frac{\boldsymbol{\overrightarrow n}}{R} \,ds\equiv\rho\,\boldsymbol{\overrightarrow n}\,ds\\ \\
	 d\boldsymbol {\overrightarrow n} = \displaystyle\left(-\frac{\boldsymbol{\overrightarrow t}}{R}+\frac{\boldsymbol{\overrightarrow b}}{\tau}  \right)\,ds\equiv \left(-\rho\,\boldsymbol {\overrightarrow t}+ \gamma\,\boldsymbol {\overrightarrow b}\right)\,ds\\ \\
	 d\boldsymbol {\overrightarrow b}=-\displaystyle\frac{\boldsymbol{\overrightarrow n}}{\tau}\,ds \equiv-\gamma\,\boldsymbol{\overrightarrow n}\,ds
\end{array}
\right. \label{FFrenet}
\end{equation}
where $R$ and $\tau$ are the radius of curvature and radius of torsion and $\rho=1/R$ and $\gamma=1/\tau$ are the  curvature and the torsion  of $(\it \Gamma)$.\\
Note that since the displacements of  unit vectors are  deformations of a solid, we can write the classical relationships
\begin{equation*}
	\left\{
	\begin{array}{l}
		d\boldsymbol {\overrightarrow t} = d\boldsymbol{\overrightarrow\omega}\times \boldsymbol {\overrightarrow t}\\ \\
		d\boldsymbol {\overrightarrow n} = d\boldsymbol{\overrightarrow \omega}\times \boldsymbol {\overrightarrow n}\\ \\
		d\boldsymbol {\overrightarrow b}=d\boldsymbol{\overrightarrow \omega}\times \boldsymbol {\overrightarrow b}\\ \\
	\end{array}
	\right. \qquad{\rm with}\quad d\boldsymbol{\overrightarrow \omega}= \boldsymbol{\overrightarrow \Omega}\,ds\quad{\rm where}\quad \boldsymbol{\overrightarrow \Omega}=\gamma\,\boldsymbol{\overrightarrow t}+\rho\,\boldsymbol{\overrightarrow n},\label{deplacement}
\end{equation*}
and \  $d\boldsymbol {\overrightarrow M} = \boldsymbol{\overrightarrow t}\, ds$.
\section{The energetic model}
\subsection {The deformation}
In reference space $\mathcal D_0$, the plant stem is represented by a small cylinder with axis $O\boldsymbol {\overrightarrow k}$, radius $r_0$ and length $L$.\\
The current point of this small cylinder is denoted $P_0$ with
\begin{equation*}
\overrightarrow{0P_0}= x_o\boldsymbol {\overrightarrow i}+y_o\boldsymbol {\overrightarrow j}+z_o\boldsymbol {\overrightarrow k}\qquad{\rm and }\qquad  \boldsymbol X_0 =\left[\ \begin{matrix}
 x_0 & \\ 
 y_0 &\\  
 z_0 
\end{matrix}\right],
\end{equation*}
 designates  the Lagrangian coordinates  in $\mathcal D_0$ \cite{Gouin_2007}.\\
In the actual space $\mathcal D$, the point $P_0$ has an image $P$ linked to the Frenet frame associated with the curve $(\it \Gamma)$ together with the point $M$ image of $M_0$ origin of the Frenet frame along curve $(\it\Gamma_0)$ (see Figure 1).
\begin{figure}[h]
	\begin{center}
		\includegraphics[width=14.2cm]{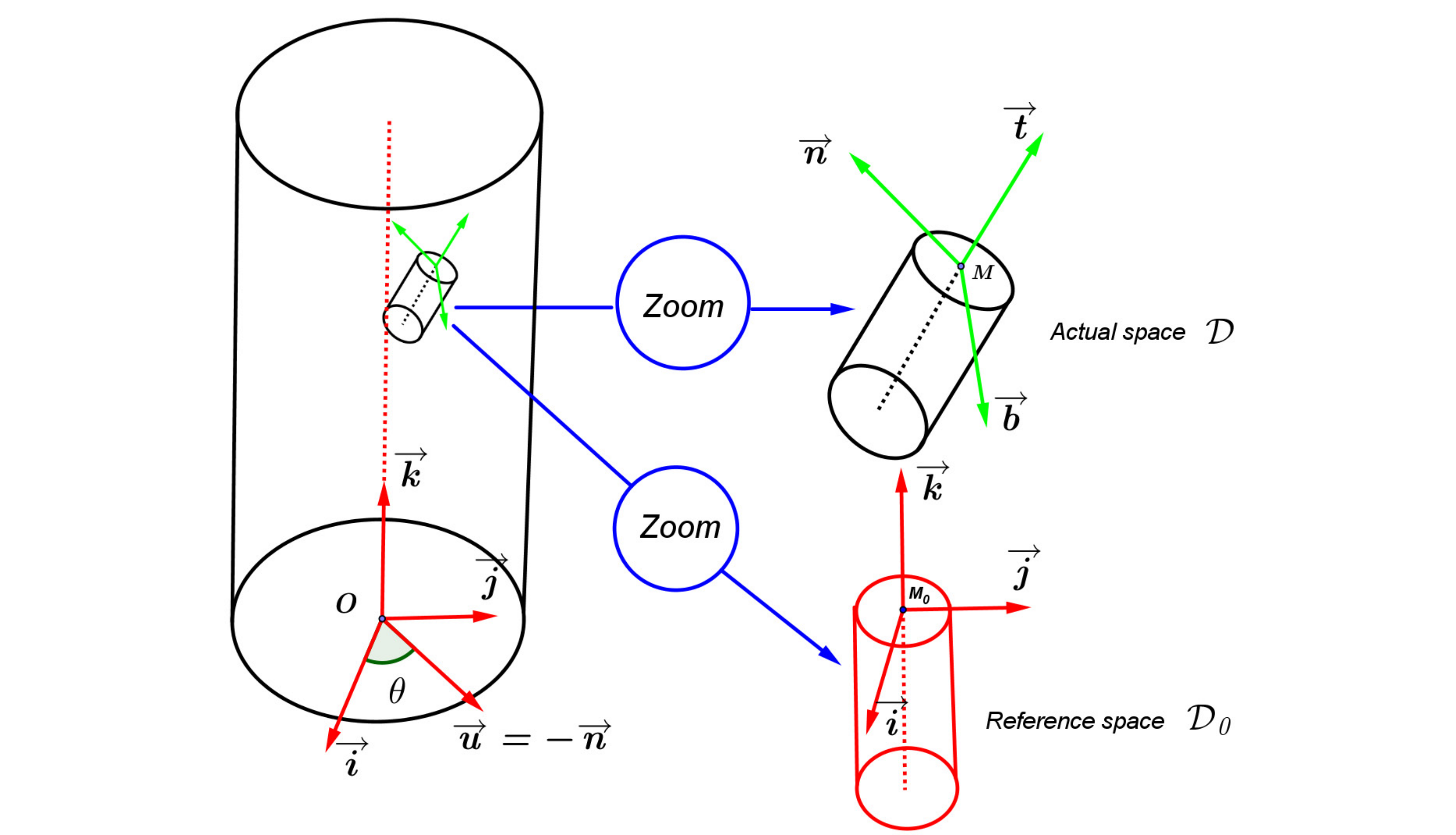}
	\end{center}
	\caption{The plant stem coils around the cylindrical stake like a helix. We have zoomed in on the positions of a fragment of the plant stem in reference space $\mathcal D_0$ and actual space $\mathcal D$.}                               
\end{figure}\\
The point $M$ on the curve $(\it\Gamma)$ of minimum length on the stake of radius $R_0$ describes a circular helix of radius $R_1=R_0+r_0$ where $r_0$, the radius of the straight section of the small cylinder, corresponds to the distance between the axis of the stake and the local axis of the plant stem (see Figure 1). We deduce that  
\begin{equation*}
{\overrightarrow {OM}} = R_1 \left(\boldsymbol {\overrightarrow u}+ a \,   \theta\,\boldsymbol{\overrightarrow k}\right)
\end{equation*}
where $\theta$ is the winding angle and $a$ denotes the pitch of the circular helix $(\it\Gamma)$, 
with, 
\begin{equation*}
	\boldsymbol{\overrightarrow u}=\cos\,\theta\,\boldsymbol{\overrightarrow i}+ \sin\, \theta\, \boldsymbol{\overrightarrow j}\quad{\rm and}\quad
	\boldsymbol{\overrightarrow v}=-\sin\,\theta\,\boldsymbol{\overrightarrow i}+ \cos\, \theta\, \boldsymbol{\overrightarrow j}.\label{coordinates}
\end{equation*}
Moreover,  
\begin{equation}
\frac{d\overrightarrow{OM}}{d\theta}=R_1\boldsymbol {\overrightarrow v}+ a\,R_1 \,\boldsymbol{\overrightarrow k}\label{dOM}.
\end{equation}
We have two possibilities, depending on   chirality of the helix and whether $\theta$ is positively or negatively oriented. In fact, the results are analytically the same, so we consider the positive orientation of $\theta$. Then, 
\begin{equation*}
ds = R_1\sqrt{1+a^2}\,d\theta\quad\Longrightarrow\quad s = R_1\sqrt{1+a^2}\,\theta\
\end{equation*}
where $s$ is the curvilinear abscissa  of $(\it \Gamma)$ positively oriented, associated with the origin, where we assume $\theta =0$.\\ By straightforward calculation, we obtain
\begin{equation}
\boldsymbol{\overrightarrow t}= \frac{\boldsymbol{\overrightarrow v}+a\,\boldsymbol{\overrightarrow k}}{\sqrt{1+a^2}},\quad \boldsymbol {\overrightarrow n} = -\boldsymbol{\overrightarrow u} ,\quad \boldsymbol{\overrightarrow b}=\frac{\boldsymbol{\overrightarrow k}-a\,\boldsymbol{\overrightarrow v}}{\sqrt{1+a^2}} \label{Frenet}
\end{equation}
and consequently from \eqref{FFrenet},
\begin{equation*}
\rho={\frac{1}{R_1\,\left(1+a^2\right)}},\quad{\rm and}\quad \gamma={\frac{a}{R_1\,\left(1+a^2\right)}}.
\end{equation*}
In this way, the $\rho$--curvature and the $\gamma$--twist remain constant along the plant stem and Frenet frame $M_0\, \boldsymbol {\overrightarrow{i}, \overrightarrow{j}, \overrightarrow{k}}$ transforms into Frenet frame $M\, \boldsymbol {\overrightarrow{t}, \overrightarrow{n}, \overrightarrow{b}}$ with no change in length or angle.  Since the length of the plant stem remains the same as in the reference space, we obtain 
\begin{equation*}
\overrightarrow{MP} = x_0\,\boldsymbol{\overrightarrow n}+y_0\,\boldsymbol{\overrightarrow b}+\left(z_0- s\right)\,\boldsymbol{\overrightarrow t}\equiv\left(z_0- R_1\sqrt{1+a^2}\,\theta\right)\,\boldsymbol{\overrightarrow t}-x_0\,\boldsymbol{\overrightarrow u}+y_0\,\boldsymbol{\overrightarrow b},
\end{equation*}
and  
\begin{equation*}
\overrightarrow{OP} = \left(R_1-x_0\right)\,\boldsymbol{\overrightarrow u}+\left(\frac{z_0-a\,y_0}{\sqrt{1+a^2}}-R_1\,\theta\right)\boldsymbol {\overrightarrow v}+\left(\frac{y_0+a\,z_0}{\sqrt{1+a^2}}\right)\boldsymbol {\overrightarrow k}.
\end{equation*}
If we write in  Eulerian coordinates  ${\overrightarrow OP} = x\,\boldsymbol{\overrightarrow i}+y\,\boldsymbol{\overrightarrow j}+z\,\boldsymbol{\overrightarrow k}$, we obtain  
\begin{equation*}
	\left\{ 
	\begin{array}{l}
		x= \left(R_1-x_0\right)\,\cos\theta- \left(\displaystyle\frac{z_0-a\,y_0}{\sqrt{1+a^2}}-R_1\,\theta\right)\sin\theta\\ \\
		y = \left(R_1-x_0\right)\,\sin\theta+  \left(\displaystyle\frac{z_0-a\,y_0}{\sqrt{1+a^2}}-R_1\,\theta\right)\cos\theta\\ \\
		z=\displaystyle\frac{y_0+a\,z_0}{\sqrt{1+a^2}}\\ \\
	\end{array}
	\right. 
\end{equation*}
and we deduce the linear tangent application $\displaystyle \frac{\partial \boldsymbol x}{\partial \boldsymbol X_0} $ associated with the deformation is
\begin{equation*}
\frac{\partial \boldsymbol x}{\partial \boldsymbol X_0} =\left[ \begin{matrix}
	\frac{\displaystyle\partial x}{\displaystyle\partial x_0} & \frac{\displaystyle\partial x}{\displaystyle\partial y_0} & \frac{\displaystyle\partial x}{\displaystyle\partial z_0}\\ \\
	\frac{\displaystyle\partial y}{\displaystyle\partial x_0} & \frac{\displaystyle\partial y}{\displaystyle\partial y_0} & \frac{\displaystyle\partial y}{\displaystyle\partial z_0}\\ \\
	\frac{\displaystyle\partial z}{\displaystyle\partial x_0} & \frac{\displaystyle\partial z}{\displaystyle\partial y_0} & \frac{\displaystyle\partial z}{\displaystyle\partial z_0}\\
\end{matrix}
\right],\quad{\rm where}\quad  
  \boldsymbol x =\left[\ \begin{matrix}
	x & \\ 
		y &\\  
		z 
	\end{matrix}\right].
\end{equation*}
Consequently, we obtain the   deformation matrix $D$ of the plant stem at each of its points which can be written as \cite{Mandel_1966} 
\begin{equation*}
	D =\left[ \begin{matrix}
		-\cos\theta&\frac{\displaystyle\sin\theta}{\displaystyle 2}\left(\displaystyle\frac{a}{\displaystyle\sqrt{1+a^2}}-1\right)& -\frac{\displaystyle\sin \theta}{\displaystyle 2\,\sqrt{1+a^2}}\\ \\
		\frac{\displaystyle\sin\theta}{\displaystyle 2}\left(\displaystyle\frac{a}{\displaystyle\sqrt{1+a^2}}-1\right) &  -	\frac{\displaystyle a\,\cos\theta}{\displaystyle \sqrt{1+a^2}} & \frac{\displaystyle 1}{\displaystyle2\,\sqrt{1+a^2}}(1+\cos\theta)\\ \\
		-\frac{\displaystyle\sin \theta}{\displaystyle 2\,\sqrt{1+a^2}} & \frac{\displaystyle 1}{\displaystyle2\,\sqrt{1+a^2}}(1+\cos\theta) & \frac{\displaystyle a}{\displaystyle \sqrt{1+a^2}}\\
	\end{matrix}
	\right] .
\end{equation*}
\\
\noindent We   deduce
\begin{equation*}
	{\rm Tr} \,D= \frac{\displaystyle a}{\displaystyle \sqrt{1+a^2}}-\cos\theta\left(1+\frac{a}{\displaystyle \sqrt{1+a^2}}\right),
\end{equation*}
where ${\rm Tr}$ denotes the trace operator and, 
\begin{equation*}
	({\rm Tr} \,D)^2 =\frac{a^2}{1+a^2}   +\cos^2\theta \left(1+\frac{\displaystyle a}{\displaystyle \sqrt{1+a^2}}\right)^2-\frac{2\,a}{\displaystyle \sqrt{1+a^2}} \left(1+\frac{a}{\displaystyle \sqrt{1+a^2}}\right)\, \cos\theta ,             
\end{equation*}
together with,
\begin{eqnarray*}
& &{\rm Tr} \,(D^2) =   \cos^2\theta \left(1+\frac{\displaystyle a^2}{\displaystyle {1+a^2}}+\frac{\displaystyle 1}{\displaystyle {2(1+a^2)}}\right)+\\ & &      \displaystyle \frac{\sin^2\theta}{2}\left(1+\frac{\displaystyle a^2}{\displaystyle {1+a^2}}-\frac{2\,a}{\displaystyle \sqrt{1+a^2}}+\frac{\displaystyle 1}{\displaystyle {(1+a^2)}}\right) 
+\frac{\cos\theta}{1+a^2}+\frac{\displaystyle 1}{\displaystyle {2(1+a^2)}} +\frac{\displaystyle a^2}{\displaystyle {1+a^2}}.         
\end{eqnarray*}
\subsection{Energy of deformation}

We denote 	$\mathcal{E_D}$
the energy per unit of volume due to  deformation of the stem. In linear elasticity we have  
\begin{equation*}
	2\,\mathcal{E_D}=\lambda\,	({\rm Tr} \,D)^2+2\,\mu \, {\rm Tr} \,(D^2),
\end{equation*}
where $\lambda$ and $\mu$ are the  Lamé coefficients of the plant stem.  We have the following results \cite{Mandel_1966}  
\begin{equation}
 E = \frac{\mu(3\lambda+2\mu)}{\lambda+\mu}, \qquad\,\ \nu =\frac{\lambda}{2(\lambda+\mu)}, 
\label{lambda-mu}
\end{equation}
where $E$ and $\nu$ are the Young modulus and Poisson  coefficient, respectively. 
The volume element of the stem is $dv =dS\times ds$ where $S$  is the area of  small cross-section   of $(\it \Gamma)$.
This volume element is an approximation for plant stems with small radii relative to the tutor radius. We have obtained
\begin{equation*}
	ds =\sqrt{1+a^2}\, R_1\, d\theta\quad {\rm where}\quad \theta \geq 0 ,\,\ {\rm and }\,\ \theta =0 \,\ {\rm corresponds}\,\ to\,\ s=0.
\end{equation*}
For calculation purposes, we consider the simplest case of an exact number of revolution coiling of the plant stem. 
For an exact number $k\in {\it\mathbb{N}}^\star$ of spiral turns of the plant stem around its stake, we obtain the length $L$  
\begin{equation*}
	L= 2\,k\pi \, R_1\,\sqrt{1+a^2}.
\end{equation*}
Then, the energy of deformation $\mathcal W_{\mathcal D}$ of the plant stem in domain $\mathcal D$ verifies  
\begin{equation*}
	2\,\mathcal W_{\mathcal D}=\iiint_{\mathcal D}\left[\lambda\,	({\rm Tr} \,D)^2+2\,\mu \, {\rm Tr} \,(D^2) \right]\, dv, \quad{\rm where}\quad  dv = R_1\,\sqrt{1+a^2}\, dS\, d\theta.
\end{equation*}
From,
\begin{equation*}
	\int_0^{2k \pi}\cos^2\theta \, d\theta = \int_0^{2k \pi}\sin^2\theta \, d\theta=k\pi\quad{\rm and}\quad \int_0^{2k \pi}\cos \theta \, d\theta=0
\end{equation*}
we obtain 
\begin{equation*}
4\,	\mathcal W_{\mathcal D}=\iint_{\it \Sigma} \left[\int_0^{2k\pi}	\left(\lambda\,({\rm Tr} \,D)^2+2\,\mu \, {\rm Tr} \,(D^2)\right)\,ds \right]\, dS,
\end{equation*}
here $(\it \Sigma)$ represents the straight cross-section of the plant stem,   $S=\pi \, r_0^2$ is the area value of $(\it \Sigma)$, and $({\rm Tr} \,D), \, {\rm Tr} \,(D^2)$ are assumed to be almost constant at each point of $(\it \Sigma)$. \\
Straightforward calculations yield   
\begin{equation*}
	\iiint_{\mathcal D}( {\rm Tr} \, D )^2\, dv = \frac{S\,L}{2}\left[1+\frac{3\,a^2}{1+a^2}+\frac{2\,a}{\displaystyle \sqrt{1+a^2}}\right]
\end{equation*}
and 
\begin{equation*}
	\iiint_{\mathcal D}{\rm Tr} \,(D^2)\, dv = \frac{S\,L}{2}\left[2+\frac{3\,a^2}{1+a^2}+\frac{3}{2\displaystyle ({1+a^2})}-\frac{a}{\displaystyle \sqrt{1+a^2}}\right].
\end{equation*}
Consequently, we obtain
\begin{equation*}
		\begin{array}{l}
\displaystyle	4\,\mathcal W_{\mathcal D}=   {S\,L} \left\{\lambda\left[1+\frac{3\,a^2}{1+a^2}+\frac{2\,a}{\displaystyle \sqrt{1+a^2}}\right]\right.\\ \displaystyle +\left. 2\,\mu\left[2+\frac{3\,a^2}{1+a^2}+\frac{3}{2\displaystyle ({1+a^2})}-\frac{a}{\displaystyle \sqrt{1+a^2}}\right]\right\}
\end{array}
\end{equation*}
\subsection{Potential energy of gravity forces}
Since the stake is a vertical cylindrical rod, the differential of the potential energy of the forces due to gravity has the expression
\begin{equation*}
	d\mathcal W_{\mathcal P}=g\,z\,dm,
\end{equation*}
where $z$ is the height of the center   of the straight cross-section of the plant stem, $g$ is the acceleration of gravity. Then,
\begin{equation*}
	z = a\,R_1\,\theta,\quad dm = \sigma_0\, S\, ds,\
\end{equation*}
where $\sigma_0$ is the volume mass (density)  assumed constant  of the plant stem. 
We get,
\begin{equation*}
	\mathcal W_{\mathcal P}= \sigma_0\, g\, S\,R_1^2\, a\,\sqrt{1+a^2}\ \frac{\theta^2}{2}, \quad{\rm with}\quad \frac{\theta^2}{2}= \frac{L^2}{\displaystyle 2\, R_1^2(1+a^2)}
\end{equation*}
or,
\begin{equation*}
	\mathcal W_{\mathcal P}= \sigma_0\, g\, S  \frac{a\,}{2\displaystyle\sqrt{1+a^2}}\, L^2.  
\end{equation*}
\subsection{Turgidity potential energy}
The action of turgidity can be estimated as that of a pressure exerted on the top of   surface   boundary of the plant stem, and which can be represented on the end of the stem by the vector  $P_{_{\mathcal T}}\,\overrightarrow{\boldsymbol{k}}$,  where $P_{_{\mathcal T}}$ denotes the turgor pressure \cite{Ali_2023,Lockhart_1965}.\\
Its work is associated with the displacement of the top of  stem. From 
\eqref{dOM} and \eqref{Frenet} we obtain
$$ {\overrightarrow {dOM}}=\left(R_1 \overrightarrow{\boldsymbol v}+ a\,R_1 \,\overrightarrow{\boldsymbol{k}}\right) d\theta = {\overrightarrow{\boldsymbol{t}}}\,ds$$  So,
\begin{equation*}
	d\mathcal T_{_{\mathcal T}}= P_{_{\mathcal T}}\, S\,\overrightarrow{\boldsymbol{k}}.\overrightarrow{\boldsymbol{t}}\, ds= P_{_{\mathcal T}}\,S  \,\frac{a}{\displaystyle \sqrt{1+a^2}}\, ds,
\end{equation*}
and one deduces the potential energy due to turgidity
\begin{equation*}
	\mathcal W_{_{\mathcal T}}= - P_{_{\mathcal T}}\,S  \,\frac{\displaystyle a}{\displaystyle \sqrt{1+a^2}}\, L
\end{equation*}
\subsection{Consequences}
The total potential energy of the plant stem is the sum of three energies of  gravity, turgidity and elastic deformation. We get
\begin{equation*}
	\begin{array} {ccc}
		2\,\mathcal W = 
	 \displaystyle \sigma_0\, g\, S  \frac{a\,}{\displaystyle\sqrt{1+a^2}}\, L^2   - 2\, P_{_{\mathcal T}}\,S  \,\frac{\displaystyle a}{\displaystyle \sqrt{1+a^2}}\, L \\ \\
	 \displaystyle +\frac{S\,L}{2}\left\{\lambda\left[1+\frac{3\,a^2}{1+a^2}+\frac{2\,a}{\displaystyle \sqrt{1+a^2}}\right]  + \mu\left[2+\frac{3\,a^2}{1+a^2}+\frac{3}{2\displaystyle ({1+a^2})}-\frac{a}{\displaystyle \sqrt{1+a^2}}\right]\right\} 
	\end{array}
\end{equation*}
or
\begin{equation*}
	\begin{array} {ccc}
\displaystyle	\frac{2\,	\mathcal W}{SL} =  
	\left(\displaystyle \sigma_0\, g\,L- 2\,P_{_{\mathcal T}}\right)\displaystyle \frac{a}{ \sqrt{1+a^2}}\\ \\
	\displaystyle +	\frac{1}{2} \left\{\lambda\left[1+\frac{3\,a^2}{1+a^2}+\frac{2\,a}{\sqrt{1+a^2}}\right] 
		+     2\,\mu\left[2+\frac{3\,a^2}{1+a^2}+\frac{3}{2\displaystyle ({1+a^2})}-\frac{a}{\displaystyle \sqrt{1+a^2}}\right]\right\} 
	\end{array}
\label{energy}
\end{equation*}
 where we recall that at the given length $L= 2\,k\pi \, R_1\,\sqrt{1+a^2}$  corresponds  an exact number   of spiral coiling of  plant stem around   its tutor.\\
For   given $R_1$,   $S$ and $L$, the mass $M= \sigma_0\,S\,L$ of the plant stem is given. The form of the plant stem corresponds to   value of  pitch  $a$ of the circular helix  associated with the minimum of energy   $\mathcal W$.

\section{Numerical application}
\begin{figure}[h]
	\begin{center}
		\includegraphics[width=12
		cm]{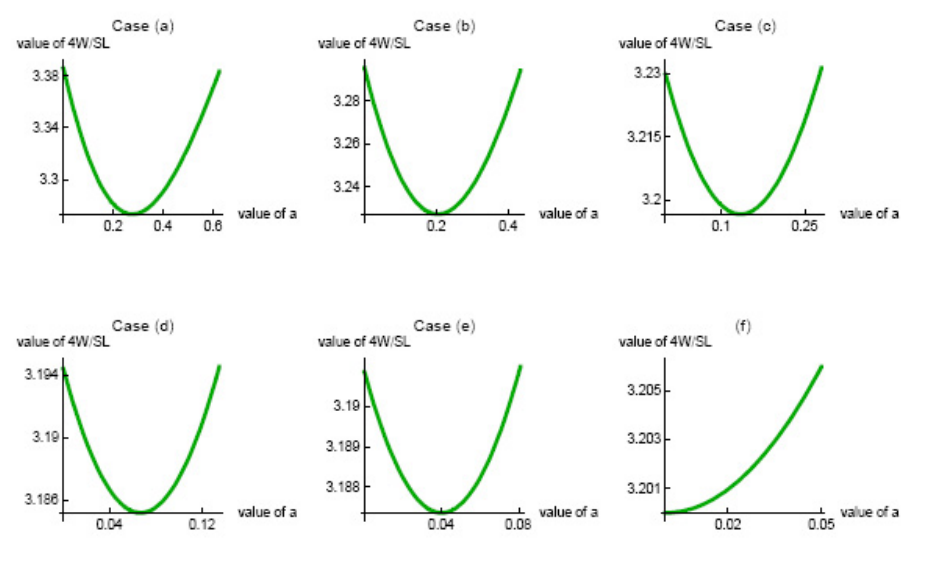}
	\end{center}
	\caption{From left to right, we have plotted the graphs for cases (a) to (f).  The x--axis represents pitch value   of the helix, the y--axis represents value   of the elastic energy per unit of volume.}\label{Fig.2}
\end{figure}
 For wood, we have $E$ of the order of $1$ to $3$ GPa ($1$ Giga Pascal = $10^9$ Pascal). For a flexible plant stem, we may estimate that $E$ is of the order of  Giga Pascal.\\
If we consider that the dimensionless Poisson's ratio is of the order of $0.05$ to $0.3$ and probably $0.05$ for a flexible plant stem, then $\lambda$ and $\mu$ given by relation \eqref{lambda-mu} are of the order of Giga Pascal ($10^9$ Pascal).\\
Terms $\sigma_0\,g \,L$ and $2\,P_{\tau}$ are of the order of $10^4$ Pascal and $10^6$ Pascal, respectively.

\begin{figure}[h]
	\begin{center}
		\includegraphics[width=9.5
		cm]{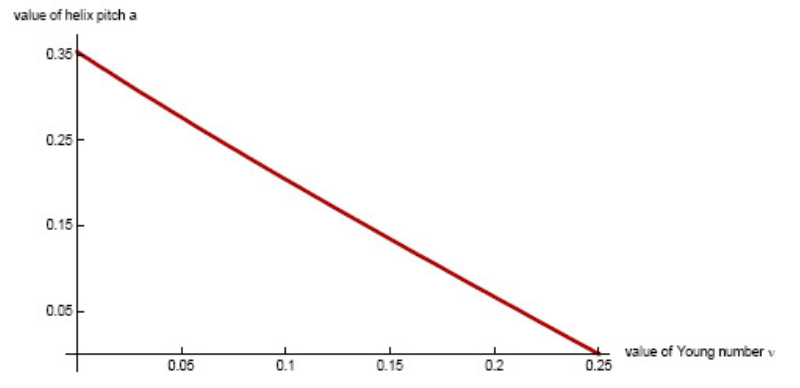}
	\end{center}
	\caption{  The x-axis plots  Poisson's number $\nu$ of elasticity in the interval  $[0, 0.25]$;  the y-axis plots the corresponding value of  pitch $a$ of  helix. }\label{Fig.3}
\end{figure} 
\noindent Consequently, \eqref{energy} can be approximated only by the terms due to the elasticity of  plant stem 
\begin{equation*}
	\frac{4\,	\mathcal W}{SL} \approx \lambda\left[1+\frac{3\,a^2}{1+a^2}+\frac{2\,a}{\displaystyle \sqrt{1+a^2}}\right]
		+ \displaystyle   2\mu\left[2+\frac{3\,a^2}{1+a^2}+\frac{3}{2\displaystyle ({1+a^2})}-\frac{a}{\displaystyle \sqrt{1+a^2}}\right]
	\label{energy1}
\end{equation*}
We have plotted the values of $\displaystyle\frac{ 4 W}{SL}$ for six cases: 
$E$ is chosen    as    reference pressure unit in \eqref{lambda-mu}$^1$ and $\nu$ is considered with different values:
\\
(a): $\nu=0.05$,\ (b): $\nu=0.1$,\ (c): $\nu=0.15$,\ (d): $\nu=0.2$,\ (e): $\nu=0.22$,\ (f): $\nu=0.25$.\\
 We obtain  graphs on Figure 2.\\
 The various graphs in Figure 2 evaluate the elastic energy per unit volume of the deformation of the plant stem. This energy admits a minimum associated with the pitch $a \in [0.05, 0.22]$ of the helix which gives the possibility of a plant stem coiling (bent and twisted) along the stake.\\ In Figure 3, we have plotted the curve for different values of helix's pitch against values of Young's modulus $\nu \in [0, 0.25]$. For approximately $\nu\succeq 0.25$ we see that the plant stem can no longer form loops along its stake.

\section{Conclusion and remarks}
For cases (a) to (e) we have obtained a minimum associated with pitch $a$ representing the set of    straight cross-sections centers of the plant stems and the corresponding height is
\begin{equation*}
z =\frac{L\,a}{\sqrt{1+a^2}}.
\end{equation*}
For case (f) the minimum is reached for zero pitch    and therefore zero height. We deduce 
\begin{equation*}
	L=\frac{V}{\pi \,r_0^2}\quad {\rm and}\quad  z =\frac{V}{\pi \,r_0^2}\, \frac{a}{\sqrt{1+a^2}}.
\end{equation*}
For a given volume $V = SL$  (where $S=\pi r_0^2$), when the    value of the radius $r_0$ of the straight cross-section of the plant stem decreases, the length $L$ increases, as does the height reached by the stem for a given pitch $a$. This height is proportional to the square of the inverse of $r_0$. Furthermore, for a stake of radius $R_0$ and $r_0$ given,    
\begin{equation*}
L =(R_0+r_0) {\sqrt{1+a^2}}\,\theta,
\end{equation*}
then, for a given length $L$, the angle $\theta$ decreases as $R_0$ increases and the number of turns of the plant stem decreases. So  larger is  the radius of the stake,   less is the plant stem wraps around its stake.\\

\noindent As we have seen, the orders of magnitude of the different energies of elasticity, gravity and turgidity show that the energy of elasticity is several orders of magnitude greater than the other two energies of gravity and turgidity. As a result, the winding of plant stems on a cylindrical stake can be generalized to plant stems that coil freely or coil on stakes of variable dimensions. We can then consider a stem element of small straight cross-section with its Frenet  frame this element being deformed relative to a reference space $\mathcal D_0$. The center of the rod element describes a $(\it\Gamma)$--curve element with a Frenet coordinate system  $M\, \boldsymbol {\overrightarrow{t}, \overrightarrow{n}, \overrightarrow{b}}$ of the current $\mathcal D$ space, deformed from the Frenet frame $M_0\, \boldsymbol {\overrightarrow{i}, \overrightarrow{j}, \overrightarrow{k}}$ of the reference space $\mathcal D_0$. The elastic energy calculations will then be analogous, and the spirals described by the $(\it\Gamma)$--curve will have deformations analogous to those obtained previously depending on  value $\nu$ of the Young modulus of elasticity of the rod.
 \bibliographystyle{plain}
 \bibliography{references} 
\end{document}